\documentclass[preprint2,twoside]{hwo}

\usepackage{graphicx}
\usepackage{enumitem}

\input{hwo.h}

\begin{document}

\title{\textbf{\LARGE Efficient, simultaneous exoplanet spectroscopy, wavefront control and calibration at deep contrast with hybrid mode-selective photonic lanterns}}
\author {\textbf{\large Mona El Morsy, $^{1}$ Olivier Guyon, $^{2,3}$ Barnaby Norris, $^{4}$ Sergio Leon-Saval, $^{5}$ Sebastien Vievard, $^{7,2}$ Julien Lozi, $^{2,3}$ Thayne Currie, $^{1,2}$, Yoo Jung Kim, $^{8}$, Mike Fitzgerald, $^{8}$ and Nemanja Jovanovic $^6$}}

\affil{$^1$\small\it Department of Physics and Astronomy, University of Texas-San Antonio, San Antonio, TX, USA}
\email{mona.elmorsy@utsa.edu}

\affil{$^2$\small\it Subaru Telescope, National Astronomical Observatory of Japan,
650 North A`oh$\bar{o}$k$\bar{u}$ Place, Hilo, HI  96720, USA}
\affil{ $^3$\small\it Astrobiology Center, 2-21-1 Osawa, Mitaka, Tokyo 181-8588, Japan}

\affil{ $^4$\small\it Sydney Institute for Astronomy, School of Physics, University of Sydney, Sydney NSW 2006, Australia}

\affil{$^5$\small\it The University of Sydney, Sydney Astrophotonic Instrumentation Laboratory, Sydney, NSW 2006, Australia}

\affil{ $^6$\small\it California Institute of Technology, Department of Astronomy, Pasadena, California 91125, USA}  

\affil{ $^7$\small\it Space Science and Engineering Initiative, College of Engineering, Institute for Astronomy, University of Hawai’i, 640 North Aohoku Place, Hilo, HI 96720, USA}

\affil{ $^8$\small\it  University of California Los Angeles, Department of Physics \& Astronomy, 475 Portola Plaza, Los Angeles, CA 90095}





\begin{abstract}
{HWO aims to directly image objects orbiting Sun-like stars, using a 6-m telescope capable of high-contrast imaging ($10^{-10}$) and spectroscopy to search for biosignatures in planets located in the habitable zone. Recent laboratory demonstrations and ground-based telescope projects have shown the effectiveness of SMFs in spectroscopy, paving the way for SMF-fed spectrographs in future space missions like HWO. SMFs enhance spectral stability and reduce modal noise. HWO spectroscopy will need extended integration times, potentially lasting weeks. During these observations, the wavefront must be precisely measured and maintained  to achieve the deep contrast and robust calibration of starlight contamination necessary for exoplanet characterization. We show that photonic lanterns (PLs) are ideally suited to meet these requirements. PLs are compact devices that couple light over a broader angular range than SMFs, ensuring higher throughput, converting a multimode input into multiple single-mode outputs. Positioned at the focal plane, they measure the complex amplitude of the coherent starlight within $\sim$ 2 l/D of the planet image, acting as compact wavefront sensors. Among the different variants of PLs that have emerged, the Hybrid-Mode Selective Photonic Lantern (HMSPL) is particularly attractive, as it directs object light into a central SMF feeding a mid-R spectrograph for exoplanet spectroscopy, while the adjacent SMFs route surrounding speckle light to a low-R spectrograph for rapid wavefront sensing. This dual function eliminates non-common path aberrations, optimizing injection efficiency and background suppression. We introduce HMSPL’s dual role and planned tests at UTSA’s high-contrast imaging lab and  at SCExAO at the Subaru Telescope.}
  \\
  \\
\end{abstract}

\section{Introduction}
Direct imaging and spectroscopy is the means by which we will someday confirm an Earth twin around a Sun-like star with a future NASA mission like HWO and a major engine of discovery about the properties of exoplanet atmospheres today \citep{CurrieBiller2023}.   While photometry and low-resolution ($R$ $\sim$ 20-100) spectroscopy are heavily utilized for direct imaging discoveries and coarse atmospheric characterization \citep[e.g.][]{Macintosh2015,Currie2023_GaiaAccelDetect}, medium resolution and especially \textit{high-resolution spectroscopy} provide an enormous wealth of information about the composition of exoplanet atmospheres \citep[e.g][]{Konopacky2013_MolecularMapping,Ruffio2021_DeepHR8799}.  Near-infrared (IR) high-resolution spectra resolve numerous molecular lines, allowing one to precisely estimate molecular abundances, constrain radial-velocities, and even inform formation environments \citep[e.g.][]{Wang2021_KPICPhaseI,Sappey2022_KPICHD206893}.   Ongoing projects such as Keck/KPIC~\citep{delorme-KPI2021}, which has been operational now for years, have spearheaded scientific advances from high-resolution spectroscopy; projects such as Subaru/REACH~\citep{kotani:20}, and VLT/HiRISE \citep{elmorsy2022, vigan:23} demonstrate the success of high-contrast, high-resolution spectroscopy on other facilities.  In all cases, these instruments' scientific capabilities leverage on an observing strategy using \textit{single mode fibers} (SMF) to sample the planet's signal.

Single Mode Fibers (SMFs) are fundamentally superior for spectroscopy requiring high measurement precision: light exiting a SMF is a stable diffraction-limited spot free of modal noise, allowing for ultra-compact spectrograph designs with a fixed line profile \citep{ghasempour:12}. 
Use of SMFs for spectroscopy is particularly promising for space mission, as their diffraction-limited PSFs free of atmospheric turbulence yields high coupling efficiency, provided that the SMF is aligned to the planet image. Coupling efficiency is then dictated by how well the fiber's propagation mode (referred to as the fundamental mode, or LP01) matches the telescope PSF \citep{coude:93,Jovanovic2017}, which is upward of 80\% for an unobstructed telescope aperture \citep{Shaklan:88} as envisioned for HWO.
Achieving high coupling efficiency is significantly harder on large ground-based telescopes due to residual atmospheric turbulence and, in some cases, large central obstructions, resulting in the source flux being distributed over a larger extent. A key innovation has been the Phononic Lantern (PL) which couples incoming light over a slightly larger angular extent to maintain high overall throughput. 

The Habitable Worlds Observatory (HWO) aims to directly image exo-Earths orbiting in the habitable zones of Sun-like stars, using a 6-meter telescope capable of $<$10$^{-10}$ contrast in reflected light and spectroscopy to search for the planets' biosignatures.   At the high spectral resolutions needed to precisely constrain biosignature abundances, 
spectroscopic observations will demand long (week(s)) integration times, during which it is essential to continuously measure and stabilize the wavefront in order to maintain a deep contrast between the exoplanet and its host star. This stabilization is crucial for ensuring accurate calibration of starlight contamination. To prevent non-common path aberrations (NCPA) that could result in uncalibrated starlight, it is necessary to monitor wavefront errors at the specific spatial frequencies corresponding to the planet's location, and at wavelengths similar to those used for science. Achieving this precise light separation is a complex challenge, especially while ensuring high observation efficiency.

\begin{figure}[ht!]
\centering
\includegraphics[width=\columnwidth]{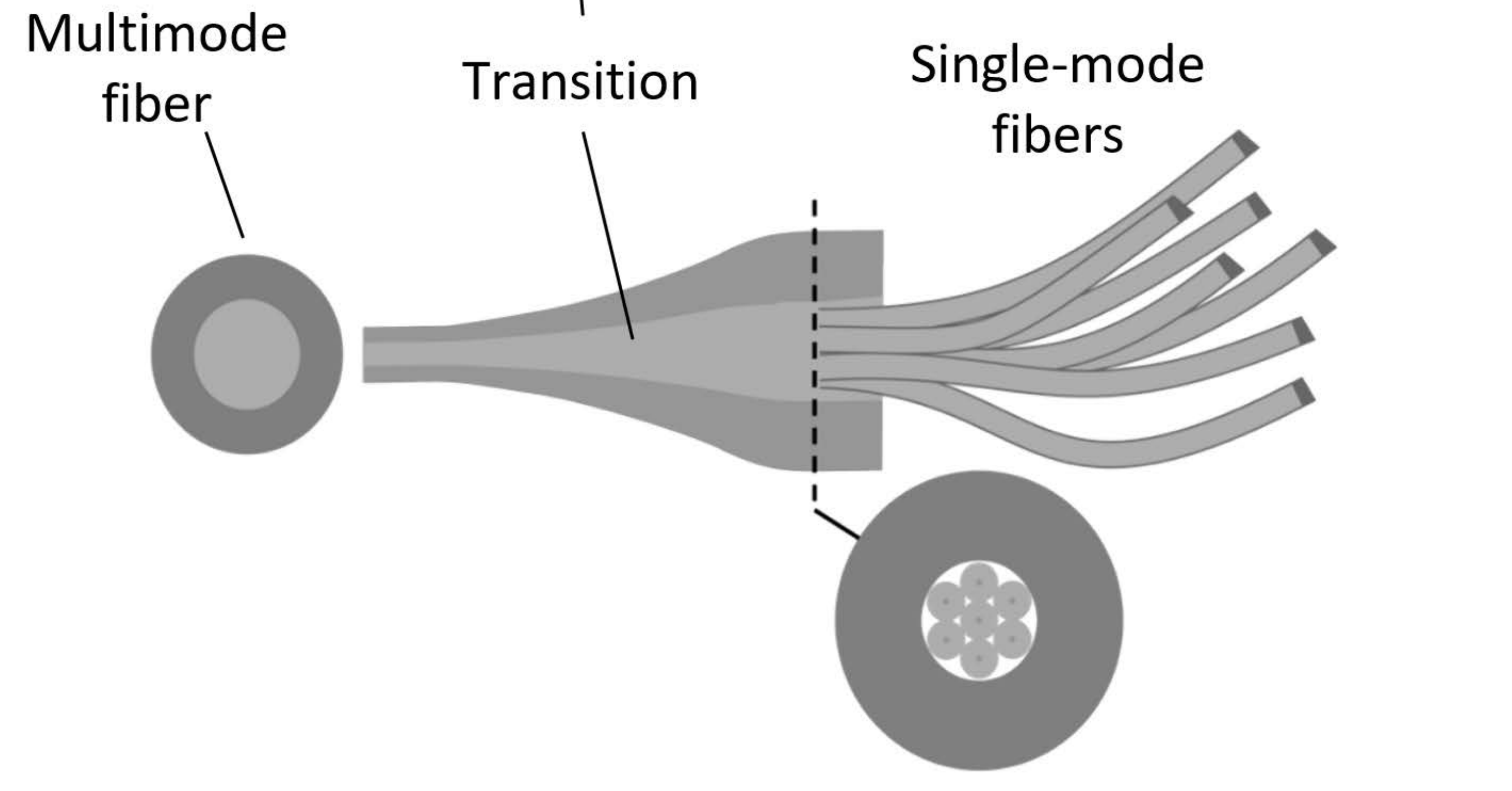}
\vspace{-0.3in}
\caption{Schematic representation of a Photonic Lantern. Light is collected by multimode fibers, and the waveguide transition to multiple single-mode fiber outputs. Adapted from \citep{Leon:10}}
\label{fig:PL}
\end{figure}
\begin{figure*}[ht!]
\centering
\includegraphics[width=\textwidth]{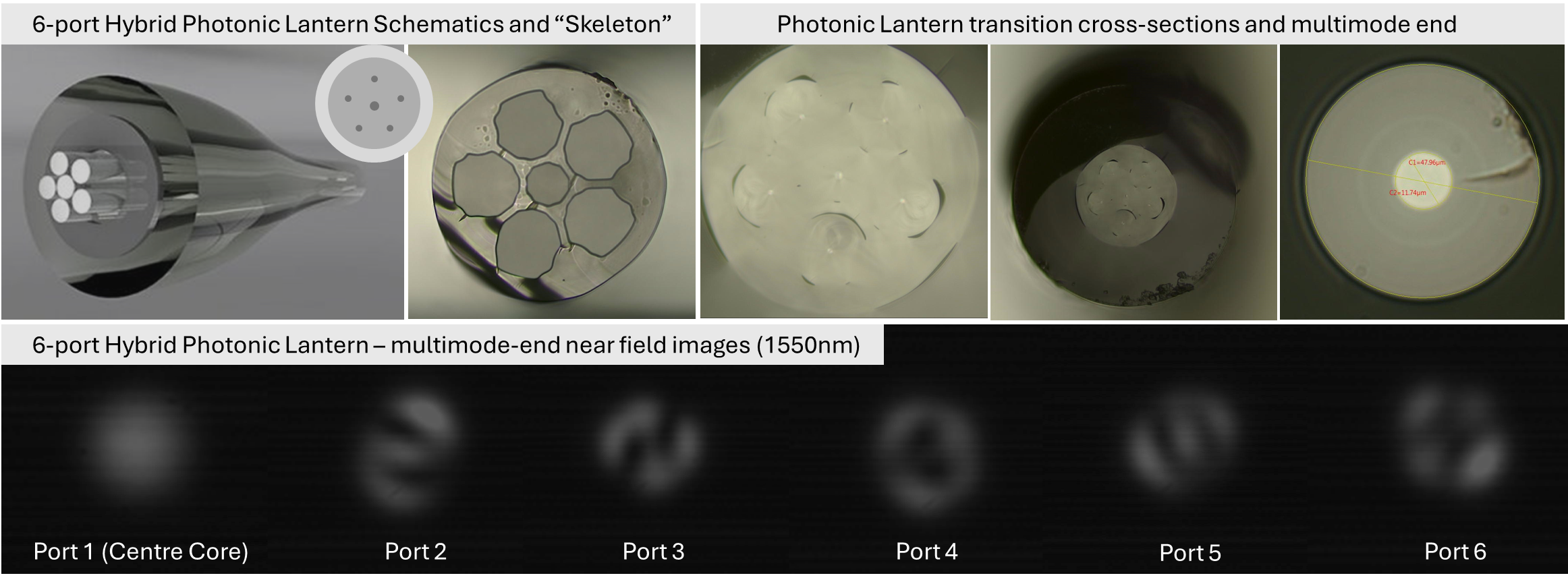}
\vspace{-0.3in}
\caption{6-port hybrid Photonic Lantern Prototype, featuring a schematic diagram, microscope images of the structure, and measured mode fields (Private communication Sergio Leon-Saval).}
\label{fig:HMS}
\end{figure*}
\section{Hybrid-Mode Selective Lantern: A promising solution to high stability, high efficiency exoplanet spectroscopy}
\subsection{Heritage of the Hybrid Mode-Selective Photonic Lantern}
The photonic lantern concept was developed to harness the coupling efficiency of multimode fibers (MMFs) while maintaining the spatial filtering advantage of single-mode fibers (SMFs). As a waveguide device, the photonic lantern converts multimode inputs into multiple SM outputs, avoiding modal noise, and focal ratio degradation (see Fig.\ref{fig:PL}). Over the years, various types have been developed, such as a photonic lantern that converts a multimode input into a 19-mode output with a throughput higher 95\% at 1.5 µm \citep{Birks:15}.

One advantage of placing photonic lanterns at the telescope’s focal plane is their ability to maintain high coupling efficiency in the presence of low-order aberrations, with their ouptut SMFs delivering diffraction-limited inputs to a downstream spectrograph. On ground-based telescope, efficient coupling still requires correcting of high order wavefront errors, provided by an upstream extreme adaptive optics (ExAO) system.

Placed at the focal plane, PLs measure the point spread function’s (PSF) complex amplitude, encoded as the relative distribution of light between the SMF ouptuts. A PL can therefore serve as a compact alternative to a traditional wavefront sensor, requiring fewer optical components and thereby minimizing non-common path aberrations (NCPA), as well as improving mechanical and thermal stability.
Photonic lanterns have recently been developed for use as focal-plane wavefront sensors \citep{lin:22}.


HMSPLs (see Fig.\ref{fig:HMS}) are designed to provide mode selectivity exclusively for the fundamental LP01 mode (where the planet light is) while mixing other modes (where WFS information is). This feature is particularly advantageous for wavefront sensing applications, where mode selectivity is not required -- in fact, optimal wavefront sensing \emph{requires} interference between sensing modes \citep{lin:22}. 
By directly mapping the LP01 mode from the input MMF to a single dedicated SMF output, HMSPLs enable precise control of static aberrations and residual low-order wavefront errors, as it performs double-duty as both wavefront-sensor and science-fiber, eliminating non-common-path aberrations.

The key challenge with HMSPLs is achieving optimal mode selectivity ($=$ exoplanet throughput to the science output SMF) across the entire H-band. The H-band is a strategic choice because the spatial filtering of the fiber helps suppress unwanted starlight, as the angular resolution of the telescope is lower in the infrared. Extensive developments in the H-band by the telecommunications and photonics industries provide a mature technological foundation that enables faster progress. To achieve a HMSPL that is optimally mode-selective for the LP01 mode across the H-band, while maintaining high throughput and sufficient outputs for wavefront sensing, a 19-port lantern design has been developed and optimized using numerical simulations.


\begin{figure*}[h!]
\centering
\includegraphics[width=\textwidth]{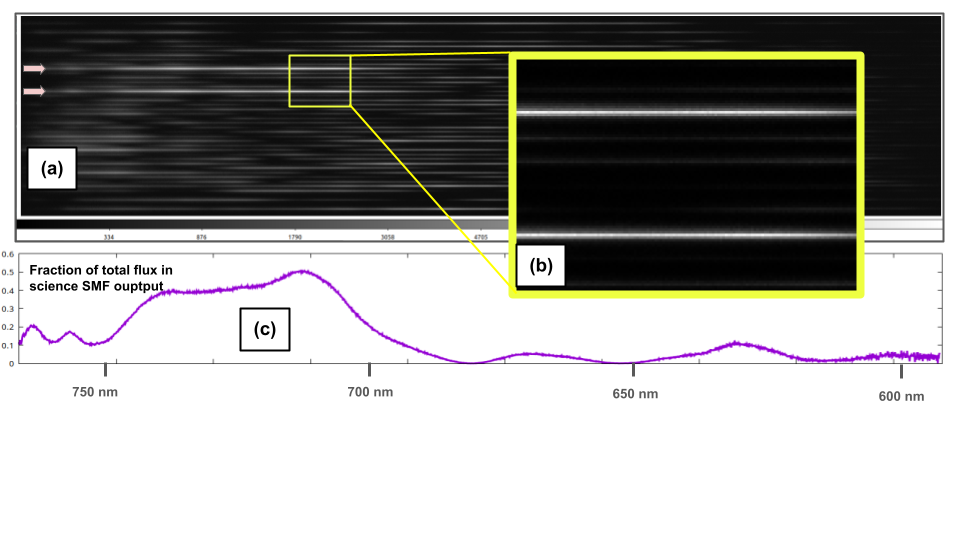}
\vspace{-2.5cm}
\caption{Output frame from the visible PL on the SCExAO instrument, acquired with SCExAO's internal source.  The raw image (a) shows 19 pairs of nearly identical spectra running horizontally. Each pair corresponds to an individual output SMF, imaged in two orthogonal polarizations. One such pair is indicated by the two light orange arrows on the left. The insert (b) shows that the spectral traces efficiently concentrate the flux on few pixels, as they are $\approx$ 1 pixel wide, and centered on detector pixel lines. This property is key to maximizing spectral resolution for faint sources. The fraction of the total light that falls in the bright SMF indicated by the light orange arrows is shown as a function of wavelength in (c). Even though this PL was not designed to be mode-selective, about 40\% of the point source light is concentrated in a single output SMF across the 710-740 nm spectral range. We will use such a configuration to start development of algorithms and measurement techniques while HMSPLs with much better flux concentration are developed (Credit: Olivier Guyon).}
\label{fig:visPLframe}
\end{figure*}

\begin{figure*}[ht!]
\centering
\includegraphics[width=\textwidth]{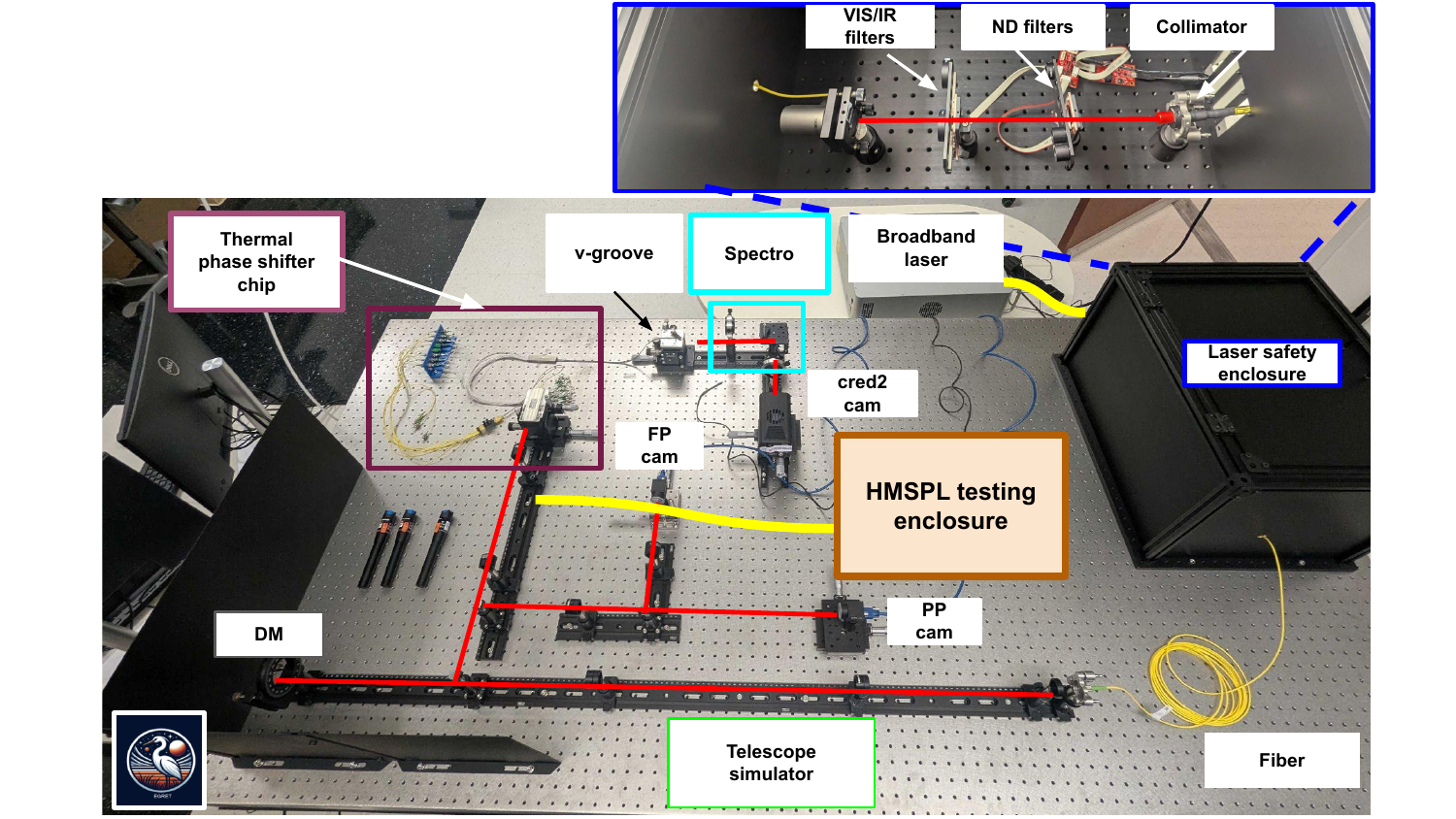}
\vspace{-0.3in}
\caption{The ExplorinG Research on Exoplanets and Technology (EGRET) testbed. A broadband source generates a light beam that passes through a laser safety enclosure, where it traverses optical filters and neutral density filters before being coupled into a fiber delivering light to the telescope simulator (lens assembly). The bench is then split into two research paths: the Thermal Phase Shifter project for GLINT \cite{norris:22}, and a fiber-coupled beam redirected to the Hybrid Mode-Selective testing enclosure (brown inset).}
\label{fig:egret}
\end{figure*}

\begin{figure}[ht!]
\centering
\includegraphics[width=\columnwidth]{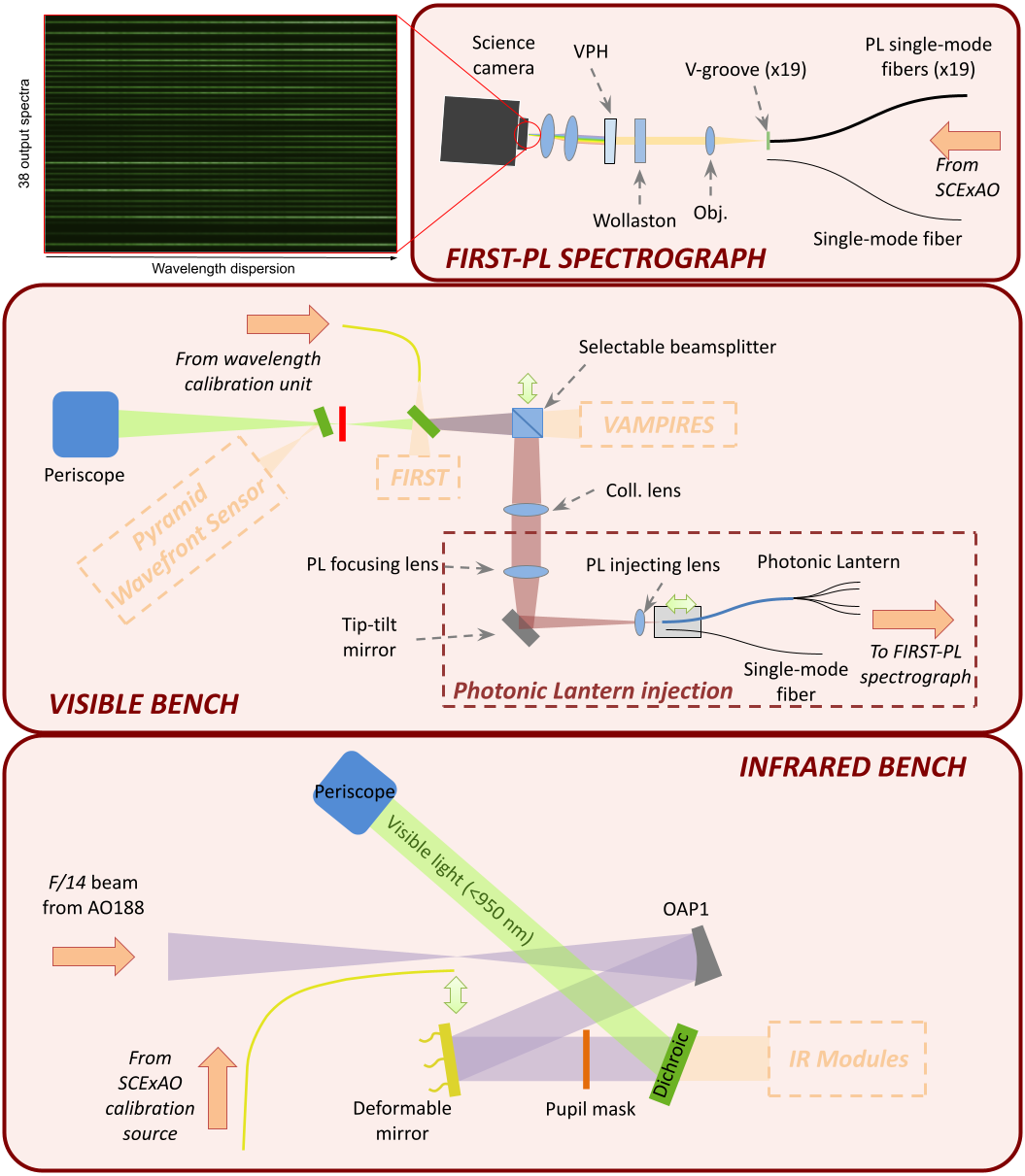}
\vspace{-0.3in}
\caption{Optical layout of SCExAO/Visible Photonic Lantern module. (Credit: Sebastien Vievard)}
\label{fig:scexao}
\end{figure}

\section{Technical Approach and Methodology}

\subsection{Overview of Laboratory Activities}

Multi-port photonic lanterns have already been manufactured and tested in laboratory and on sky in the IR and visible \citep{Xin2022_PLVFN, vievard:24}. However, efficiently coupling light into the MSPL poses challenges due to wavefront aberrations, PSF shape, and stability. The long-term stability of the photonic lantern is particularly critical, as spectroscopy observations on HWO can span weeks, during which the HMSPL would also play a key role in starlight suppression.

To address these challenges, we will implement the photonic lantern as part of UTSA's ExplorinG Research on Exoplanets and Technology (EGRET) testbed (See Fig. \ref{fig:egret}). This configuration will allow us to study the spectral response of the HMSPL over time and under varying thermal conditions, thereby providing valuable insights into its long-term stability and performance. 

\subsection{HMSPL long term stability testing at UTSA}
The stability of the HMSPL's transfer function is central to its efficacy as a high contrast imaging instrument on HWO. Optimal science-light injection and speckle rejection relies on maintaining a closed control loop between the PL (in its role as a wavefront sensor) and telescope deformable mirror and active optics, which in turn relies on the transfer function remaining constant. The PL wavefront sensor system is intrinsically more stable than conventional wavefront sensing systems, due to the fact that the entire sensing-portion of the system consists of a monolithic photonic component tens of millimeters in size, making it mechanically robust, and exposed to far smaller gradients in temperature than a traditional bulk-optical wavefront sensor. The laboratory experiments will quantify this stability. 

To date, this has not been properly measured, since the effect has proven to be smaller than the expected optical path difference changes from a standard laboratory's bulk optical components. The HWO application will require the PL transfer function to remain stable over the $\sim$week-long observation time, and in this project the requirement is for the relative PL output intensities to vary by less than 0.1\% over this time period. Once the change in transfer function is measured as a function of temperature (and for all wavelengths), both self-calibration algorithms and physical PL design can be optimized accordingly.

To perform these measurements, the test apparatus (See Fig. \ref{fig:egret}) will be placed in a modestly thermally stabilized enclosure (brown inset in Fig.\ref{fig:egret}) to maintain temperature to within $\pm$0.1~K, and then temperature sensors placed throughout the enclosure on key components will be used to determine the precise temperatures during the test run. Both residual variations in ambient temperature and active modulation of the temperature will be used to probe the transfer function's temperature response. 

During test runs, a constant PSF produced from a broadband supercontinuum light source, covering both the visible (VIS) and near-infrared (NIR) spectral ranges will be feeding the optical testbed. The source delivers an average power exceeding 3W, with a pulse duration of less than 10 ps and exceptional stability (standard deviation $<$0.5\%). The collimated output beam first passes through a reflective neutral density filter and a set of filters mounted on an automated wheel. The visible spectral band is used for bench alignment, while the near-infrared band is selected for experiments. 
The attenuated and filtered light is then injected into a SMF, mounted on a 3-axis stage and filtered by a circular aperture placed at the pupil plane. A lens system collimates the beam and directs it toward a reflective Deformable Mirror (DM) with a pupil diameter of 2.3 mm. The DM, a gold-coated HEX-111 model from Boston Micromachines, consists of 37 hexagonal segments, each 750 µm in size, with a 3.5 µm stroke and a frequency of 2 kHz.
A bulk of optics, including a beam splitter, will be positioned just upstream of the PL to redirect part of the light toward a near-IR camera (First Light CRED-2), allowing imaging of the PL entrance.
At the output of the PL, a combination of lenses and a prism will enable us to image the spectral response of the PL on a second CRED-2 camera.
To probe the entire space of the response function, tests will be performed using different input PSFs created by the testbed's DM, in order to ensure excitation of all modes occurs over the experiment set. 

\subsection{Simultaneous spectroscopy and wavefront control validation at SCExAO}

The spectroscopy and wavefront control activities will be carried out primarily using the SCExAO instrument at the Subaru Telescope (See Fig.\ref{fig:scexao}). SCExAO’s system-level capabilities complement UTSA’s EGRET testbed, which focuses on long-term stability measurements.

Initial tests at SCExAO will use the existing visible photonic lantern (PL) in a quasi–mode-selective configuration to maximize coupling into the science fiber. These will later transition to the near-IR following delivery of the NIR HMSPL. The existing SMF spectrographs cover visible and near-IR ranges at moderate spectral resolution.

SCExAO’s high-speed synchronization allows accurate characterization of the PL response to upstream wavefront changes. This will be used to identify and maintain the wavefront state and source position that optimize broadband throughput into the science fiber. “Scenes” with star, planet, and background will be emulated by combining exposures taken with different input positions.


Wavefront control tests will adapt existing control-loop methods to an off-axis planet-like source, probing performance over a range of speckle halo contrasts. These experiments will assess achievable stability, spectral resolution requirements for sensing, control loop dynamic range, and the potential for the PL to assist with active nulling.

\subsection{End-to-End on-sky validation} 
All near-infrared (NIR) data will be acquired with the existing photonic spectrograph, which already incorporates a photonic lantern backend. A system-level end-to-end demonstration will be carried out on-sky at the Subaru Telescope by targeting a bright star with a moderate-contrast ($10^{-6}$) companion at small angular separation. The HMSPL will be positioned on the companion, and observations will be conducted both with and without stellar suppression.

This test will address several aspects of the technique that cannot be reproduced under laboratory conditions. On-sky operation naturally combines the starlight, companion, and background signals without the need for synthetic scene construction. It will also provide the first demonstration of spectral extraction from the science fiber using astrophysical sources, in contrast to the featureless calibration sources used in the lab. Furthermore, the telescope environment introduces a broader and more realistic range of wavefront perturbations—including amplitude errors, telescope-induced polarization effects, and vibrations—that cannot be simulated on the bench. The full acquisition sequence, from telescope pointing to companion alignment, will also be tested in operational conditions.

Active wavefront control will be performed as best effort basis, as the rapid atmospheric variations may exceed the linearity of the wavefront sensor (WFS). High-speed WFS measurements (kHz cadence) will be obtained with the photon-counting C-Red One camera (First Light), enabling exploration of self-calibration of the extracted spectra using the WFS signals.

\section{Summary}
The Astro2020 decadal survey prioritized the development of the Habitable Worlds Observatory, with the mission of detecting and characterizing Exo-Earths around Sun-like stars. To achieve this, the survey emphasized the need for sustained research efforts in enabling technologies, with NASA setting a TRL 5 requirement by 2029 for mission-critical advancements. The astrophysics technology report (Rivera, 2024) formalized these directives by identifying key technology gaps that must be addressed in the coming decade.
The work proposed here aligns with these objectives by advancing photonic instrumentation for high-contrast imaging and wavefront sensing, which are critical capabilities for exoplanet characterization. 

The HMSPL device offers an ultra-compact, high-throughput solution that is well suited for space-based spectroscopy, enabling precise mode mapping across broadband wavelengths.
Furthermore, the proposed development of active wavefront control and calibration techniques ensures robust performance in space environments, directly supporting the high-contrast imaging requirements of HWO. Through iterative design, fabrication, laboratory tests and on-sky system validation, this effort contributes to NASA’s roadmap for equipping next-generation observatories for exoplanet characterizations.

A key outcome of this work will be to help maximize HWO's spectral resolution, which is essential for exoplanet characterization and search for biomarkers. The flight instrument resolution will be driven by multiple considerations in addition to the spectrograph efficiency, including the detector characteristics (dark current, cosmic rays). The instrument may ultimately feature several modes (low/high resolution) to adapt to the planet flux. Our laboratory results, obtained at resolutions ranging from a few 100s to few 1000s, can be extrapolated to higher/lower resolutions to help guide this important choice.

\vspace{1cm}
{\bf Acknowledgements}

The development of SCExAO was supported by the Japan Society for the Promotion of Science (Grant-in-Aid for Research \#23340051, \#26220704, \#23103002, \#19H00703 \& \#19H00695), the Astrobiology Center of the National Institutes of Natural Sciences, Japan, the Mt Cuba Foundation and the director's contingency fund at Subaru Telescope.
The authors wish to recognize and acknowledge the very significant cultural role and reverence that the summit of Maunakea has always had within the indigenous Hawaiian community. We are most fortunate to have the opportunity to conduct observations from this mountain.
\bibliography{references.bib}
\end{document}